\documentclass[journal]{IEEEtran}

\ifCLASSINFOpdf
\else
   \usepackage[dvips]{graphicx}
\fi
\usepackage{url}

\usepackage{amsmath,amsfonts}

\usepackage[caption=false,font=normalsize,labelfont=sf,textfont=sf]{subfig}

\usepackage{stfloats}
\usepackage{verbatim}

\usepackage{booktabs}
\usepackage{balance}
\usepackage{empheq, epsfig,amssymb,amsthm,url}
\usepackage{algorithm}
\usepackage{algpseudocode}
\usepackage{multirow}
\usepackage{mdframed}
\usepackage{url}
\usepackage{enumitem}
\usepackage[makeroom]{cancel}
\usepackage{array}
\usepackage{comment}
\usepackage{epstopdf}
\usepackage{float}
\usepackage{breqn}
\usepackage{cases}
\usepackage{tabularx}
\usepackage[printonlyused,withpage]{acronym}
\usepackage{balance}
\usepackage{graphbox} 

\usepackage{hyperref}
\usepackage[table]{xcolor} 
\usepackage{colortbl}
\definecolor{gray}{rgb}{0.9,0.9,0.9} 
\usepackage{tikz}
\usetikzlibrary{fit,calc}

\colorlet{mypink}{red!40}
\colorlet{myblue}{cyan!60}

\usepackage{setspace}
\setlist[itemize]{label=$\triangleright$}
\newtheoremstyle{break}
{}
{}
{\itshape}
{}
{\bfseries}
{.}
{\newline}
{}
\theoremstyle{break}

\theoremstyle{definition}



\newcommand{\vect}[1]{\mathbf{#1}}
\newcommand{\bs}[1]{\boldsymbol{#1}}
\newcommand{\E}{\mathbb{E}}

\newcommand{\il}{\texttt{IL}}
\newcommand{\tl}{\texttt{TL}}

\usepackage{url}

\makeatletter
\def\thmhead@plain#1#2#3{%
	\thmname{#1}\thmnumber{\@ifnotempty{#1}{ }\@upn{#2}}%
	\thmnote{ {\the\thm@notefont#3}}}
\let\thmhead\thmhead@plain
\makeatother

\graphicspath{{./figs/}}

\newcommand{\argmin}{\operatornamewithlimits{argmin}}

\newcommand{\lk}{ \left\{ }
\newcommand{\rk}{ \right\} }

\newcommand{\diag}{\mbox{{diag}}}

\newcommand{\Ib}{{\bf I}}

\newsavebox\mybox

\acrodef{SE}{speech enhancement}
\acrodef{AVSE}{audio-visual speech enhancement}
\acrodef{STFT}{short-time Fourier transform}
\acrodef{ESTOI}{extended short-time objective intelligibility}
\acrodef{NMF}{non-negative matrix factorization}
\acrodef{DNN}{deep neural network}
\acrodef{VAE}{variational auto-encoder}
\acrodef{DKF}{deep Kalman filter}
\acrodefplural{VAEs}{variational auto-encoders}
\acrodef{EM}{expectation-maximization}
\acrodef{TF}{time-frequency}
\acrodef{ELBO}{evidence lower bound}
\acrodef{LR}{Living Room}
\acrodef{SDR}{signal-to-distortion ratio}
\acrodef{PESQ}{perceptual evaluation of speech quality}
\acrodef{SNR}{signal-to-noise ratio}
\acrodef{DNNs}{deep neural networks}
\acrodef{VESDE}{variance-exploding stochastic differential equation}
\acrodef{SDE}{stochastic differential equation}
\acrodef{GAN}{generative adversarial networks}
\acrodefplural{GANs}{generative adversarial networks}
\acrodef{SI-SDR}{scale-invariant signal-to-distortion ratio}
\acrodef{MOS}{mean opinion score}
\acrodef{SGMSE+}{score-based generative model for speech enhancement}
\acrodef{NCSNPP++}{Noise-Conditional Score Network}
\acrodef{WSJ}{Wall Street Journal}
\acrodef{UDiffSE}{Unsupervised Diffusion-Based Speech Enhancement}
\acrodef{DIFFUSER}{Diffusion-based Fast Framework for Unsupervised Speech Enhancement}
\acrodef{PC}{Predictor-Corrector}
\acrodef{DMPS}{Diffusion Model Posterior Sampling}
\acrodef{NN}{neural network}

\newcommand{\normalc}{\mathcal{N}_{\mathbb{C}}(\vect{0}, \vect{I})}

\usepackage{bm}

\usepackage{cite}
\usepackage{textcomp}
\def\BibTeX{{\rm B\kern-.05em{\sc i\kern-.025em b}\kern-.08em
    T\kern-.1667em\lower.7ex\hbox{E}\kern-.125emX}}

\begin{document}

\setlength{\abovedisplayskip}{3.5pt}
\setlength{\belowdisplayskip}{3.5pt}

\title{Posterior Transition Modeling for Unsupervised Diffusion-Based Speech Enhancement}

\author{Mostafa Sadeghi, Jean-Eudes Ayilo, Romain Serizel, and Xavier Alameda-Pineda, \IEEEmembership{Senior Member, IEEE}
\thanks{M. Sadeghi, J.-E. Ayilo, and R. Serizel are with the Multispeech team, Université de Lorraine, CNRS, Inria, Loria, Nancy, France.}
\thanks{X. Alameda-Pineda is with the RobotLearn team, Université Grenoble Alpes, Inria, Grenoble, France.}
\thanks{This work was supported by the French National Research Agency (ANR) under the project REAVISE (ANR-22-CE23-0026-01).}
}

\markboth{Journal of \LaTeX\ Class Files, Vol. 14, No. 8, August 2015}
{Shell \MakeLowercase{\textit{et al.}}: Bare Demo of IEEEtran.cls for IEEE Journals}
\maketitle

\begin{abstract}
We explore unsupervised speech enhancement using diffusion models as expressive generative priors for clean speech. Existing approaches guide the reverse diffusion process using noisy speech through an approximate, noise-perturbed likelihood score, combined with the unconditional score via a trade-off hyperparameter. In this work, we propose two alternative algorithms that directly model the conditional reverse transition distribution of diffusion states. The first method integrates the diffusion prior with the observation model in a principled way, removing the need for hyperparameter tuning. The second defines a diffusion process over the noisy speech itself, yielding a fully tractable and exact likelihood score.  Experiments on the WSJ0-QUT and VoiceBank-DEMAND datasets demonstrate improved enhancement metrics and greater robustness to domain shifts compared to both supervised and unsupervised baselines.
\end{abstract}


\IEEEpeerreviewmaketitle

\section{Introduction}

\IEEEPARstart{D}{iffusion} models have recently emerged as a powerful class of generative models, achieving impressive results in image, audio, and video generation~\cite{song2021scorebased, croitoru2023diffusion}. These models learn expressive generative priors for complex natural data by gradually adding noise through a forward process, then training a reverse denoising process to remove it. Beyond their success in generation tasks, diffusion models have been extended to solve inverse problems \cite{daras2024survey}, owing to their ability to perform both unconditional and conditional data generation.

In the field of speech processing, diffusion models have been successfully leveraged for speech enhancement (SE), an inverse problem aimed at recovering clean speech from a noisy recording \cite{richter2023speech, lemercier2023storm, gonzalez2024diffusion, serra2022universal, lu2022conditional, yen2023cold}. {While SE can encompass tasks such as dereverberation, we focus on single-channel speech denoising}. Traditionally, deep learning–based SE has been dominated by predictive methods, which map noisy speech directly to clean speech via supervised training with paired data. Examples include regression-based convolutional or recurrent architectures that optimize mean-squared error, time-frequency masking, or perceptual loss functions \cite{wang2018supervised, xu2014regression, luo2019conv}. In contrast, diffusion-based SE incorporates a prior distribution over clean speech with a generative training objective, rather than directly learning a noisy-to-clean mapping. 

Two main approaches exist for applying diffusion models to SE. \textit{Supervised} methods learn the conditional distribution of clean speech given noisy input~\cite{richter2023speech, gonzalez2024diffusion, serra2022universal, lu2022conditional, yen2023cold}, enabling reverse diffusion conditioned on the noisy signal. In contrast, \textit{unsupervised} methods train only on clean speech to learn a prior~\cite{nortier2023unsupervised, ayilo2024diffavse, iashchenko23_interspeech, moliner2024buddy}, combining it at test time with a noise model via Bayes’ rule. Inference alternates posterior-guided denoising steps with noise parameter updates. Compared to supervised approaches, this unsupervised setting allows for reusing the trained prior for other speech inverse problems and may provide robustness to noise domain mismatch~\cite{nortier2023unsupervised, ayilo2024diffavse, lemercier2025diffusion}.

While the unsupervised framework is well studied in image processing~\cite{chung2023diffusion, daras2024survey}, its application to SE remains limited~\cite{nortier2023unsupervised, ayilo2024diffavse}. In existing methods, the posterior score is obtained by combining the learned prior score with the gradient of a noisy-speech likelihood. However, the likelihood term is time-dependent and intractable, typically approximated under simplified assumptions such as an uninformative speech prior~\cite{meng2024diffusion}. Its influence is further controlled by a trade-off hyperparameter. Moreover, these methods update only the mean of the prior (unconditional) reverse diffusion process using the posterior score, while keeping the variance fixed, which can lead to inaccurate posterior modeling and requires tuning of the guidance strength.

In this paper, we propose a novel unsupervised SE framework that performs posterior sampling by explicitly modeling the conditional (reverse) transition distribution of the diffusion states. Our first algorithm integrates the diffusion prior and observation model in a principled way, eliminating the need for a guidance hyperparameter. The second algorithm extends this approach by introducing a diffusion process over the noisy speech itself, enabling exact computation of the time-dependent likelihood. We evaluate both methods on the WSJ0-QUT and VoiceBank-DEMAND benchmarks and observe consistent improvements (mainly with the second algorithm) over existing unsupervised approaches across standard enhancement metrics. Notably, the second algorithm exhibits robustness to domain shifts between training and test conditions, often matching or surpassing supervised baselines.

\section{Background and related work}\label{sec:udiffse}
We review the diffusion-based unsupervised {SE} framework \cite{nortier2023unsupervised, ayilo2024diffavse}, which forms the foundation of our approach. Speech signals are first represented in the \ac{STFT} domain, resulting in a complex-valued 2D array of shape $F \times L$ ($F$ frequency bins, $L$ time frames). For notational convenience, this array is vectorized into $\vect{s} \in \mathbb{C}^{FL}$. Speech enhancement is performed using a prior distribution over clean speech signals learned by a diffusion model. This prior is defined via a continuous-time diffusion process $\{\vect{s}_{t}\}_{t\in[0, T]}$, which gradually corrupts clean data $\vect{s}_0 = \vect{s}$ into Gaussian noise as the diffusion time $t$ increases (not to be confused with the time index of the STFT representation). The corruption is governed by a forward \ac{SDE} \cite{song2021scorebased, nortier2023unsupervised}:
\begin{equation}\label{eqn:sde-fwd}
    \mathsf{d}\vect{s}_t = \vect{f}(\vect{s}_t) \mathsf{d}t + g_t \mathsf{d}\vect{w},
\end{equation}
where $\vect{f}(\vect{s}_t)=-\gamma\vect{s}_t$ is the drift coefficient ($\gamma>0$), $g_t$ controls noise variance, and $\vect{w}$ is a standard Wiener process. {The solution of \eqref{eqn:sde-fwd} defines a diffusion process whose conditional marginal distributions are characterized by the perturbation kernel}
$p_{0t}(\vect{s}_t | \vect{s}) = \mathcal{N}_{\mathbb{C}}(\mathsf{e}^{-\gamma t} \vect{s}, \sigma_t^2\vect{I})$, where $\sigma_t^2$ is determined from the \ac{SDE}. Under mild conditions~\cite{anderson1982}, this process is reversible via a reverse \ac{SDE}:
\begin{equation}\label{eqn:rev-sde}
    \mathsf{d}\vect{s}_t = [\vect{f}(\vect{s}_t) - g_t^2 \nabla_{\vect{s}_t} \log p_t(\vect{s}_t) ] \mathsf{d}t + g_t \mathsf{d}\vect{\bar{w}},
\end{equation}
where $\vect{\bar{w}}$ is a backward Wiener process, and $\nabla_{\vect{s}_t} \log p_t(\vect{s}_t)$ (the score function) is approximated by a \ac{DNN}-based score model $\vect{S}_{\theta}(\vect{s}_{t}, {t})$. The model parameters are learned by solving the following problem \cite{kingma2023understanding}:
\begin{equation}\label{eqn:train-obj}
    \theta^{*} = {\argmin_{\theta} \mathbb{E}_{t, \vect{s}, \bs{\zeta}, \vect{s}_t | \vect{s}}}
    {\Bigr[\| {\sigma_t}{\vect{S}_{\theta}(\vect{s}_{t}, {t})}+ {\bs{\zeta}} \|_2^2 \Bigl]},
\end{equation}
where $\bs{\zeta}\sim\mathcal{N}_{\mathbb{C}}(\vect{0}, \vect{I})$ represents zero-mean Gaussian noise. The trained score model ${\vect{S}_{\theta^*}(\vect{s}_{t}, {t})}$ then replaces the score function in \eqref{eqn:rev-sde}, and the resulting reverse \ac{SDE} is solved numerically to generate clean speech samples starting from Gaussian noise.

To perform speech enhancement, the noisy speech signal is modeled as $\vect{x} = \vect{s} + \vect{n}$, where $\vect{n} \sim \mathcal{N}_{\mathbb{C}}(\bs{0}, \diag(\vect{v}_{\phi}))$ represents complex Gaussian noise with structured variance $\vect{v}_{\phi} = \text{vec}(\vect{W}\vect{H})$, parameterized by low-rank non-negative matrices $\vect{W}, \vect{H}$. The parameters $\phi = \{\vect{W}, \vect{H}\}$ are optimized using the \ac{EM} algorithm \cite{bishop2006pattern}. The M-step maximizes the expected log-likelihood $\E_{p(\vect{s}|\vect{x})}\lk\log p_\phi(\vect{x}| \vect{s})\rk$, with the expectation approximated by Monte Carlo sampling from the posterior $p(\vect{s}|\vect{x}) \propto p_{\phi}(\vect{x}|\vect{s}) p(\vect{s})$ in the E-step. 

Posterior sampling is performed by replacing the unconditional score in \eqref{eqn:rev-sde} with a conditional one: $\nabla_{\vect{s}_t} \log p(\vect{s}_t|\vect{x}) = \nabla_{\vect{s}_t} \log p_{\phi}(\vect{x}|\vect{s}_t)+ \nabla_{\vect{s}_t} \log p_t(\vect{s}_t)$. The second term is approximated by the trained score model $\vect{S}_{\theta^*}(\vect{s}_{t}, {t})$. Since $p_{\phi}(\vect{x}|\vect{s}_t)$ is intractable for $t>0$, it is approximated under an uninformative prior assumption on $\vect{s}$ \cite{nortier2023unsupervised}, yielding
\begin{equation}\label{eqn:psllkd}
    \tilde{p}_\phi(\vect{x}|\vect{s}_t) \sim \mathcal{N}_{\mathbb{C}}\Big(\mathsf{e}^{\gamma t}\vect{s}_t,\, {\sigma_t^2}\mathsf{e}^{2\gamma t} \vect{I} + \diag(\vect{v}_{\phi})\Big).
\end{equation}
The resulting conditional reverse SDE is then expressed as 
\begin{equation}\label{eqn:post-rev-sde}
        \mathsf{d}\vect{s}_t  =  \Big[\vect{f}(\vect{s}_t) - g_t^2 \Big(\vect{S}_{\theta^*}(\vect{s}_{t}, {t})+ \lambda_t \nabla_{\vect{s}_t} \log \tilde{p}_\phi (\vect{x}|\vect{s}_t)\Big) \Big] \mathsf{d}t + g_t \mathsf{d}\vect{\bar{w}},
\end{equation}
where $\lambda_t>0$ balances the likelihood's impact. This gradient guidance approach enables the reverse process to generate speech signals that are consistent with $\vect{x}$. The above reverse \ac{SDE} is numerically solved to estimate clean speech for the current value of $\phi$, which is then used to update $\phi$. The EM steps are repeated until convergence, typically within five iterations \cite{nortier2023unsupervised}. A more efficient algorithm was recently proposed~\cite{ayilo2024diffavse}, in which $\phi$ is updated at each reverse iteration using an intermediate speech estimate derived from Tweedie's formula \cite{efron2011tweedie}:
\begin{equation}\label{eq:tweedie}
    \hat{\vect{s}}_{0,t}=
    \E_{p_{t 0}(\vect{s}|\vect{s}_t)}[\vect{s}]\approx\frac{\vect{s}_t + \sigma_{t}^2 \vect{S}_{\theta^*}(\vect{s}_t, t)}{\mathsf{e}^{-\gamma t}}.
\end{equation}
This method avoids a full reverse \ac{SDE} process at each step to generate speech samples for updating $\phi$, significantly improving computational efficiency. These unsupervised approaches stand in contrast to supervised frameworks~\cite{richter2023speech, lemercier2023storm}, which directly learn the conditional score $\nabla_{\vect{s}_t} \log p(\vect{s}_t|\vect{x})$ during training from paired clean and noisy speech data.

\section{Proposed framework}\label{sec:proposed}
We present \texttt{DEPSE} (Diffusion-based Explicit Posterior Sampling for Speech Enhancement), which integrates the diffusion prior and the observation model by deriving the conditional transition distribution. {The proposed framework concerns only the E-step of the EM algorithm, while the M-step remains identical to that used in prior work \cite{ayilo2024diffavse}.} Section~\ref{sec:prior_sampling} revisits diffusion-based prior sampling as used in~\cite{nortier2023unsupervised, ayilo2024diffavse}. Section~\ref{sec:depse-v1} introduces \texttt{DEPSE-IL}, which removes the need for the balancing hyperparameter~$\lambda_t$ while still approximating the intractable likelihood (IL). Section~\ref{sec:depse-v2} presents \texttt{DEPSE-TL}, which defines a diffusion SDE over noisy speech, resulting in a tractable likelihood (TL).

\vspace{-2mm}
\subsection{Diffusion prior sampling formulation}\label{sec:prior_sampling}
We adopt the Euler–Maruyama method as a numerical solver for the reverse SDE \eqref{eqn:rev-sde} with time discretization. This yields a finite sequence of latent variables, denoted by $\{ \vect{s}_i \}_{i=0}^{N}$, which form a Markov chain $\vect{s}_0\rightarrow \vect{s}_1 \rightarrow\ldots \rightarrow \vect{s}_{N}$, with
\begin{equation}\label{eq:joint_prior}
    p(\vect{s}_0,\ldots,\vect{s}_{N})=p(\vect{s}_{N})\prod_{i=1}^N   p(\vect{s}_{{i-1}}|\vect{s}_{i}).
\end{equation}
Discretizing \eqref{eqn:rev-sde} and inserting the score model leads to:
\begin{equation}\label{eq:disc-rev}
    {\vect{s}}_{{i-1}} = {\vect{s}}_{i} - \vect{f}_i\Delta{\tau} + g_{\tau_i}^2 \vect{S}_{\theta^*}(\vect{s}_i, \tau_i)\Delta{\tau}+  g_{\tau_i}\sqrt{\Delta{\tau}}\bs{\zeta},
\end{equation}
with $\vect{s}_{N}\sim \mathcal{N}_{\mathbb{C}}(\bs{0}, \Ib)$. Here, $\{ \tau_1,\ldots,\tau_N\}$ is an equally spaced sequence of time steps in $[0,T]$ with a discretization width of $\Delta{\tau}=T/N$, $\bs{\zeta}\sim\mathcal{N}_{\mathbb{C}}(\bs{0}, \Ib)$, and $\vect{f}_i=-\gamma \vect{s}_i$. By applying a similar discretization to the forward SDE in \eqref{eqn:sde-fwd}, we obtain:
\begin{equation}\label{eqn:si_s0}
    p(\vect{s}_i | \vect{s}_0) = \mathcal{N}_{\mathbb{C}}(\mathsf{e}^{-\gamma\tau_{i}} \vect{s}_0, \sigma_{\tau_{i}}^2\vect{I}), \quad i\ge 1.
\end{equation}
Since we have access to an approximate score function, we can improve the numerical SDE solver by incorporating a score-based Markov Chain Monte Carlo (MCMC) method, such as Langevin MCMC, used as a corrector step in prior works~\cite{song2021scorebased, nortier2023unsupervised, ayilo2024diffavse}. Combined with \eqref{eq:disc-rev}, unconditional sampling is then characterized by a backward transition distribution:
\begin{equation}\label{eq:cond_prior}
    p(\vect{s}_{{i-1}}|\vect{s}_{i})=\mathcal{N}_{\mathbb{C}}\Big(\bs{\mu}^{\mathsf{back}}(\vect{s}_{i}), \bs{\Sigma}^{\mathsf{back}}_{i}\Big),
\end{equation}
with
\begin{equation}\label{eq:mu_sigma_update}
  \left\{
  \mathopen{}\begin{array}{l}
    \bs{\mu}^{\mathsf{back}}(\vect{s}_{i}) =\vect{h}({\vect{s}}_{i}) - \vect{f}_{i}\Delta{\tau} + g_{\tau_i}^2 \vect{S}_{\theta^*}(\vect{h}({\vect{s}}_{i}), \tau_i)\Delta{\tau} \\[2mm]
    \bs{\Sigma}^{\mathsf{back}}_{i} = g_{\tau_i}^2\Delta{\tau}\Ib.
  \end{array}
  \right.
\end{equation}
where $\vect{h}({\vect{s}}_{i})$ is the result of one-step Langevin MCMC: 
\begin{equation}
    \vect{h}({\vect{s}}_{i}) = \vect{s}_{i} + \epsilon_{\tau_i} {\vect{S}_{\theta^*}(\vect{s}_{i}, \tau_i)} + \sqrt{2\epsilon_{\tau_i}}\bs{\zeta},\quad \bs{\zeta}\sim\normalc.
\end{equation}
Here, $\epsilon_{\tau_i} = (\sigma_{\tau_i} \cdot r)^2$ denotes the Langevin step size ($r>0$) \cite{song2022solving}. Starting from Gaussian noise, i.e., $\vect{s}_{N}\sim \mathcal{N}_{\mathbb{C}}(\bs{0}, \Ib)$, the above conditional sampling procedure is iterated to ultimately generate a speech signal $\vect{s}_0$ from the learned distribution.  

\vspace{-2mm}
\subsection{Explicit posterior sampling for speech enhancement}\label{sec:depse-v1}
While diffusion prior sampling effectively captures the distribution of clean speech $\vect{s}$, it does not incorporate any information about the noisy observation $\vect{x}$. Therefore, we target the conditional distribution for SE, expressed as follows:
\begin{equation}\label{eq:joint_post}
    p_\phi(\vect{s}_0,\ldots,\vect{s}_{N}|\vect{x} )=p(\vect{s}_{N})\prod_{i=1}^N   p_\phi(\vect{s}_{{i-1}}|\vect{s}_{i}, \vect{x}).
\end{equation}
We aim to approximate the conditional transition distribution $p_\phi(\vect{s}_{{i-1}}|\vect{s}_{i}, \vect{x})$ given the current $\phi$, which is expressed as
\begin{algorithm}[t!]
   \small
  \setstretch{1.4}  
\caption{\texttt{DEPSE-IL}}\label{alg:diffuser_v1}
\begin{algorithmic}[1]
    \State \textbf{Input:} $\vect{x}$, $\{ \tau_1,\ldots,\tau_N\}$
    \State ${\vect{s}}_N \sim \mathcal{N}_{\mathbb{C}}(\vect{x}, \sigma^2_{\tau_N}\vect{I})$
    \For{$i = N,\ldots, 1 $}
        \State \text{Update} $\{\bs{\mu}^{\mathsf{back}}(\vect{s}_{i}), \bs{\Sigma}^{\mathsf{back}}_{i}\}$ using \eqref{eq:mu_sigma_update}
        \State \text{Update} $\{\bs{\mu}^{\il}_\phi(\vect{s}_{i}, \vect{x}), \bs{\Sigma}^{\il}_{s, i}(\phi)\}$ using \eqref{eq:post_mu_sigma_update1}
        \State $\vect{s}_{{i-1}} = \bs{\mu}^{\il}_\phi(\vect{s}_{i}, \vect{x}) + \sqrt{\bs{\Sigma}^{\il}_{s, i}(\phi)}\bs{\zeta}, \quad \bs{\zeta}\sim \mathcal{N}_{\mathbb{C}}(\bs{0},\, \Ib)$
        \State $\hat{\vect{s}}_{0,i} = (\vect{s}_{i}+\sigma_{\tau_i}^2\vect{S}_{\theta^*}(\vect{s}_i, \tau_i)) / \mathsf{e}^{-\gamma\tau_{i}} $
        \State {$\phi = \{ \vect{W}, \vect{H} \} \gets \mathsf{NMF}(| \vect{x} - \hat{\vect{s}}_{0,i} |^2)$}
            
    \EndFor
    \State \textbf{Output:} $\hat{\vect{s}}={\vect{s}}_{0}$
\end{algorithmic}
\end{algorithm}
\begin{equation}\label{eq:cond_posterior2}
    p_\phi(\vect{s}_{{i-1}}|\vect{s}_{i}, \vect{x})\propto p_\phi(\vect{x}|\vect{s}_{{i-1}})\, p(\vect{s}_{{i-1}}|\vect{s}_{i}).
\end{equation}
The second term is provided by the prior transition distribution in \eqref{eq:cond_prior}, and for the likelihood term, we write
\begin{equation}\label{eq:likelihood_x2}
    p_\phi(\vect{x}|\vect{s}_{{i-1}}) = \int 
p_\phi(\vect{x}|\vect{s}_0)\,p(\vect{s}_0|\vect{s}_{{i-1}})\,\mathsf{d}\vect{s}_0.
\end{equation}
Similar to \cite{nortier2023unsupervised, ayilo2024diffavse}, we make an uninformative prior assumption for $\vect{s}_0$, leading to $p(\vect{s}_0|\vect{s}_{{i-1}})\propto p(\vect{s}_{{i-1}}|\vect{s}_0)$. Combined with \eqref{eqn:si_s0}, we obtain:
\begin{equation}\label{eq:likelihood_approx}
    p_\phi(\vect{x}|\vect{s}_{{i-1}})\approx \mathcal{N}_{\mathbb{C}}\Big(\mathsf{e}^{\gamma\tau_{i-1}}\vect{s}_{i-1},\, \bs{\Sigma}_{x, i}^{\il}(\phi)\Big),
\end{equation}
where $\bs{\Sigma}_{x, i}^{\il}(\phi)
 = \frac{\sigma_{\tau_{i-1}}^2}{\mathsf{e}^{-2\gamma\tau_{i-1}}} \Ib + \diag(\vect{v}_{\phi})$. The conditional transition distribution then takes the following form 
\begin{equation}\label{eq:depse_v1_post}
    p_\phi(\vect{s}_{{i-1}}|\vect{s}_{i}, \vect{x})=\mathcal{N}_{\mathbb{C}}\Big(\bs{\mu}^{\il}_\phi(\vect{s}_{i}, \vect{x}),\, \bs{\Sigma}^{\il}_{s, i}(\phi)\Big)
\end{equation}
with
\begin{equation}\label{eq:post_mu_sigma_update1}
\small
  \left\{
  \mathopen{}\begin{array}{l}
    \displaystyle \bs{\mu}^{\il}_\phi(\vect{s}_{i}, \vect{x}) \;=\; 
\bs{\Sigma}^{\il}_{s, i}(\phi)
\biggl( 
  \frac{\bs{\mu}^{\mathsf{back}}(\vect{s}_{i})}{\bs{\Sigma}^{\mathsf{back}}_{i}} 
  \;+\; 
  \frac{\vect{x}}{\mathsf{e}^{-\gamma\tau_{i-1}}\bs{\Sigma}_{x, i}^{\il}(\phi)}
\biggr), \\[3mm]
    \displaystyle \bs{\Sigma}^{\il}_{s, i}(\phi) \;=\; 
\frac{
  \mathsf{e}^{-2\gamma\tau_{i-1}}\bs{\Sigma}_{x, i}^{\il}(\phi)\,\bs{\Sigma}^{\mathsf{back}}_{i}
}{
  \mathsf{e}^{-2\gamma\tau_{i-1}}\bs{\Sigma}_{x, i}^{\il}(\phi)
  \;+\;  
  \bs{\Sigma}^{\mathsf{back}}_{i}
}.
  \end{array}
  \right.
\end{equation}
Speech enhancement is performed through iterative sampling from the conditional transition distribution \eqref{eq:depse_v1_post}, with the parameter $\phi$ updated at each step using the Tweedie-based speech estimate given in \eqref{eq:tweedie}. The complete \texttt{DEPSE-IL} procedure is summarized in Algorithm~\ref{alg:diffuser_v1}. In contrast to prior methods~\cite{nortier2023unsupervised, ayilo2024diffavse}, which perform posterior sampling by modifying only the mean of the reverse transition distribution~\eqref{eq:mu_sigma_update} using a posterior score in~\eqref{eqn:post-rev-sde} while keeping the variance fixed, our method explicitly derives and samples from the conditional transition distribution~\eqref{eq:depse_v1_post}. This leads to more accurate posterior modeling and eliminates
the need for the balancing hyperparameter $\lambda_t$, whose tuning is often non-trivial~\cite{UDiffSE2024Supplement}.
\begin{algorithm}[t!]
   \small
  \setstretch{1.4}  
\caption{\texttt{DEPSE-TL}}\label{alg:diffuser_v2}
\begin{algorithmic}[1]
    \State \textbf{Input:} $\vect{x}$, $\{ \tau_1,\ldots,\tau_N\}$
    \State ${\vect{s}}_N \sim \mathcal{N}_{\mathbb{C}}(\vect{x}_N, \sigma^2_{\tau_N}\vect{I})$
    \For{$i = N,\ldots, 1 $}
        \State \text{Update} $\{ \bs{\mu}^{\mathsf{back}}(\vect{s}_{i}), \bs{\Sigma}^{\mathsf{back}}_i\}$ using \eqref{eq:mu_sigma_update}
        \State $\vect{x}_{{i-1}} = \mathsf{e}^{-\gamma\tau_{i-1}}\vect{x} + \sigma_{\tau_{i-1}}\bs{\zeta}, \quad \bs{\zeta}\sim \mathcal{N}_{\mathbb{C}}(\bs{0},\, \Ib)$
        \State \text{Update} $\{\bs{\mu}^{\tl}_\phi(\vect{s}_{i}, \vect{x}_{{i-1}}), \bs{\Sigma}^{\tl}_{s, i}(\phi)\}$ using \eqref{eq:post_mu_sigma_update2}
        \State $\vect{s}_{{i-1}} = \bs{\mu}^{\tl}_\phi(\vect{s}_{i}, \vect{x}_{{i-1}}) + \sqrt{\bs{\Sigma}^{\tl}_{s, i}(\phi)}\bs{\zeta}, \quad \bs{\zeta}\sim \mathcal{N}_{\mathbb{C}}(\bs{0},\, \Ib)$
        \State $\hat{\vect{s}}_{0,i} = (\vect{s}_{i}+\sigma_{\tau_i}^2\vect{S}_{\theta^*}(\vect{s}_i, \tau_i)) / \mathsf{e}^{-\gamma\tau_{i}} $
        \State {$\phi = \{ \vect{W}, \vect{H} \} \gets \mathsf{NMF}(| \vect{x} - \hat{\vect{s}}_{0,i} |^2)$}
            
    \EndFor
    \State \textbf{Output:} $\hat{\vect{s}}={\vect{s}}_{0}$
\end{algorithmic}
\end{algorithm}

\vspace{-2mm}
\subsection{Incorporating noisy speech diffusion process}\label{sec:depse-v2}
We extend \texttt{DEPSE-IL} by introducing a diffusion process over the noisy speech signal $\vect{x}$, inspired by \cite{song2022solving, song2021scorebased, dou2024diffusion}, {which define a measurement-space diffusion to simplify likelihood computation in the reverse process}. {Unlike \texttt{DEPSE-IL} and earlier methods \cite{nortier2023unsupervised, ayilo2024diffavse}, which approximate the time-dependent likelihood with an uninformative prior assumption (see \eqref{eqn:psllkd}, \eqref{eq:likelihood_approx}),} this second approach, \texttt{DEPSE-TL}, yields a tractable likelihood and avoids such approximations. {We define the set of latent variables $\{ \vect{x}_i\}_{i=0}^N$ ($\vect{x}_0 = \vect{x}$), following the same forward SDE as in \eqref{eqn:sde-fwd}. Similar to \eqref{eqn:si_s0}, we can write $p(\vect{x}_{i-1}|\vect{x})=\mathcal{N}_{\mathbb{C}}( \mathsf{e}^{-\gamma\tau_{i-1}}\vect{x},\, \sigma^2_{\tau_{i-1}}\Ib)$.} {Given the observation model $\vect{x} = \vect{s} + \vect{n}$, the time-dependent likelihood distribution can be computed in closed-form:}
\begin{equation}\label{eq:exact_likelihood}
    p_\phi(\vect{x}_{{i-1}}|\vect{s}_{{i-1}}) = \mathcal{N}_{\mathbb{C}}\Big(\vect{s}_{{i-1}},\, \bs{\Sigma}_{x, i}^{\tl}(\phi) \Big),
\end{equation}
with $\bs{\Sigma}_{x, i}^{\tl}(\phi) = {\mathsf{e}^{-2\gamma\tau_{i-1}}} \diag(\vect{v}_{\phi})$. The corresponding posterior distribution takes the following form:
\begin{equation}\label{eq:cond_posterior3}
    p_\phi(\vect{s}_{{i-1}}|\vect{s}_{i}, \vect{x}_{i-1})\propto p_\phi(\vect{x}_{i-1}|\vect{s}_{{i-1}})\, p(\vect{s}_{{i-1}}|\vect{s}_{i}).
\end{equation}
Substituting the expressions for the two distributions on the right, the conditional transition distribution is derived as:
\begin{equation}\label{eq:final_cond_post2}
    p_\phi(\vect{s}_{{i-1}}|\vect{s}_{i}, \vect{x}_{i-1})=\mathcal{N}_{\mathbb{C}}\Big(\bs{\mu}^{\tl}_\phi(\vect{s}_{i}, \vect{x}_{i-1}),\, \bs{\Sigma}^{\tl}_{s, i}(\phi)\Big),
\end{equation}
where the mean and covariance are given by:
\begin{equation}\label{eq:post_mu_sigma_update2}
\small
  \left\{
  \mathopen{}\begin{array}{l}
    \displaystyle \bs{\mu}^{\tl}_\phi(\vect{s}_{i}, \vect{x}_{i-1}) \;=\; 
\bs{\Sigma}^{\tl}_{s, i}(\phi)
\biggl( 
  \frac{\bs{\mu}^{\mathsf{back}}(\vect{s}_{i})}{\bs{\Sigma}^{\mathsf{back}}_{i}} 
  \;+\; 
  \frac{\vect{x}_{i-1}}{\bs{\Sigma}_{x, i}^{\tl}(\phi)}
\biggr), \\[3mm]
    \displaystyle \bs{\Sigma}^{\tl}_{s, i}(\phi) \;=\; 
\frac{
  \bs{\Sigma}_{x, i}^{\tl}(\phi)\,\bs{\Sigma}^{\mathsf{back}}_{i}
}{
  \bs{\Sigma}_{x, i}^{\tl}(\phi) 
  \;+\;  
  \bs{\Sigma}^{\mathsf{back}}_{i}
}.
  \end{array}
  \right.
\end{equation}
Algorithm~\ref{alg:diffuser_v2} summarizes the overall \texttt{DEPSE-TL} algorithm.

\section{Experiments}\label{sec:exp}
\noindent\textbf{Baselines.} We compare the proposed algorithms with UDiffSE+ (unsupervised-generative)~\cite{ayilo2024diffavse}, Conv-TasNet (supervised, adapted for SE {and trained with the SNR objective})~\cite{luo2019conv}, and SGMSE+ (supervised-generative)~\cite{richter2023speech}. {These are not state-of-the-art (SOTA), as we focus on robustness under domain mismatch rather than SOTA performance.}

\noindent\textbf{Model architectures.} For the unsupervised and SGMSE+ algorithms, we use a lightweight variant of the Noise Conditional Score Network (NCSN++) \cite{richter2023speech}, with 5.2~M parameters. All networks, including Conv-TasNet, are trained from scratch.

\noindent\textbf{Hyperparameter setting.} Speech data are sampled at 16~kHz. STFT is computed using a Hann window of size 510 with 75\% overlap. SDE hyperparameters follow \cite{nortier2023unsupervised}. The reverse SDE is discretized into $N=30$ steps. In UDiffSE+, we set $\lambda_{\tau_i} = 1.5$ for even $i$, and $0$ otherwise, as suggested in \cite{ayilo2024diffavse}.

\noindent\textbf{Dataset.} We use WSJ0-QUT \cite{leglaive2020recurrent} and VoiceBank-DEMAND (VB-DMD) \cite{botinhao2016investigating}. WSJ0-QUT is constructed by mixing clean speech from the Wall Street Journal corpus (WSJ0) \cite{garofolo1993csr} with noise from QUT-Noise \cite{dean2015qut} (e.g., \textit{car}, \textit{café}, \textit{street}, \textit{living room}, \textit{kitchen}). VB-DMD combines clean speech from the Voice Bank corpus (VB) with noise from the DEMAND dataset \cite{thiemann2013diverse} (e.g., \textit{living room}, \textit{office space}, \textit{bus}, \textit{open-area cafeteria}, \textit{public square}, \textit{restaurant}, \textit{subway}). SE models are trained either on the WSJ0-QUT training set (25 hours) or on the VB-DMD corpus (9.39 hours), and evaluated on the test sets of WSJ0-QUT (1.48 hours) or VB-DMD test set (0.58 hour). Unsupervised models are trained only on the corresponding clean speech data.
Noisy speech signals in WSJ0-QUT are generated with SNRs of \{-5, 0, 5\} dB (for both training and testing). In VB-DMD, SNRs are \{0, 5, 10, 15\} dB for training and \{2.5, 7.5, 12.5, 17.5\} dB for testing.

\noindent\textbf{Evaluation metrics.} We report standard SE metrics: scale-invariant signal-to-distortion, interference, and artifact ratios (SI-SDR, SI-SIR, SI-SAR) in dB \cite{le2019sdr}—{though these are not ideal for evaluating generative models}—as well as the extended short-term objective intelligibility (ESTOI) \cite{jensen2016algorithm} ($[0,1]$) and the perceptual evaluation of speech quality (PESQ) score \cite{Rix2001pesq} ($[-0.5,4.5]$). Higher values indicate better performance.

\begin{table}[t]
\captionsetup{font=footnotesize}
\caption{Average results on WSJ0-QUT under matched (WSJ0-QUT) and mismatched (VB-DMD) training conditions (\textbf{overall best}, \underline{unsupervised best}). ``Sup.'' indicates supervision type: S = supervised, U = unsupervised.}
\setlength{\tabcolsep}{5pt} 
\centering
\scriptsize
\begin{tabular}{@{}clcccccc@{}} 
\toprule
 & Method & Sup. & SI-SDR & SI-SIR & SI-SAR & PESQ & ESTOI \\ 
\midrule
\multirow{8}{*}{\rotatebox[origin=c]{90}{Matched}} 
& Input & -- & -2.60 & -2.60 & 46.66 & 1.83 & 0.50 \\
\midrule
& SGMSE+ \cite{richter2023speech} & S & 8.53 & 24.65 & 8.72 & 2.61 & 0.76 \\
& Conv-TasNet \cite{luo2019conv} & S & \textbf{12.63} & \textbf{29.88} & \textbf{12.75} & \textbf{2.86} & \textbf{0.83} \\
& UDiffSE+ \cite{ayilo2024diffavse} & U & 3.34 & \underline{12.34} & 4.34 & \underline{2.19} & 0.58 \\
& \texttt{DEPSE-IL} & U & 3.41 & 10.61 & 4.83 & \underline{2.19} & 0.59 \\
& \texttt{DEPSE-TL} & U & \underline{3.89} & 6.55 & \underline{7.93} & 2.17 & \underline{0.61} \\
\midrule
\multirow{5}{*}{\rotatebox[origin=c]{90}{Mismatched}} 
& SGMSE+ \cite{richter2023speech} & S & 3.02 & 12.79 & 3.86 & 2.08 & 0.61 \\
& Conv-TasNet \cite{luo2019conv} & S & \textbf{5.83} & \textbf{17.07} & \textbf{6.36} & \textbf{2.21} & \textbf{0.63} \\
& UDiffSE+ \cite{ayilo2024diffavse} & U & 0.06 & \underline{8.72} & 1.21 & 2.05 & 0.52 \\
& \texttt{DEPSE-IL} & U & 0.18 & 6.55 & 1.86 & 2.05 & 0.53 \\
& \texttt{DEPSE-TL} & U & \underline{2.66} & 5.96 & \underline{{6.06}} & \underline{2.17} & \underline{0.59} \\
\bottomrule
\end{tabular}
\label{tab:wsj}
\end{table}
\begin{table}[t]
\captionsetup{font=footnotesize}
\caption{Average results on VB-DMD under matched (VB-DMD) and mismatched (WSJ0-QUT) training conditions (\textbf{overall best}, \underline{unsupervised best}).  ``Sup.'' indicates supervision type: S = supervised, U = unsupervised.}
\setlength{\tabcolsep}{5pt} 
\centering
\scriptsize
\begin{tabular}{@{}clcccccc@{}} 
\toprule
 & Method & Sup. & SI-SDR & SI-SIR & SI-SAR & PESQ & ESTOI \\ 
\midrule
\multirow{8}{*}{\rotatebox[origin=c]{90}{Matched}} 
& Input & -- & 8.45 & 8.45 & 47.51 & 3.02 & 0.79 \\
\midrule
& SGMSE+ \cite{richter2023speech} & S & 17.16 & 28.89 & 17.65 & \textbf{3.51} & \textbf{0.85} \\
& Conv-TasNet \cite{luo2019conv} & S & \textbf{19.26} & \textbf{32.36} & \textbf{19.80} & 3.40 & \textbf{0.85} \\
& UDiffSE+ \cite{ayilo2024diffavse} & U & 10.40 & \underline{22.45} & 11.04 & 3.30 & 0.77 \\
& \texttt{DEPSE-IL} & U & 10.49 & 19.73 & 11.54 & 3.36 & 0.79 \\
& \texttt{DEPSE-TL} & U & \underline{13.87} & 16.40 & \underline{18.38} & \underline{3.41} & \underline{0.82} \\
\midrule
\multirow{5}{*}{\rotatebox[origin=c]{90}{Mismatched}} 
& SGMSE+ \cite{richter2023speech} & S & 10.06 & 12.43 & \textbf{17.13} & 3.29 & 0.82 \\
& Conv-TasNet \cite{luo2019conv} & S & 11.13 & 15.22 & 15.90 & 3.28 & \textbf{0.83} \\
& UDiffSE+ \cite{ayilo2024diffavse} & U & 10.87 & \underline{\textbf{28.27}} & 11.02 & 3.30 & 0.77 \\
& \texttt{DEPSE-IL} & U & 11.06 & 24.58 & 11.41 & 3.31 & 0.78 \\
& \texttt{DEPSE-TL} & U & \underline{\textbf{14.01}} & 17.56 & \underline{\textbf{17.37}} & \underline{\textbf{3.34}} & \underline{0.81} \\
\bottomrule
\end{tabular}
\label{tab:vb}
\end{table}
\noindent\textbf{Results.} 
Tables \ref{tab:wsj} and \ref{tab:vb} report the average metric values for evaluation on the WSJ0-QUT and VB-DMD test sets, respectively, both with matched and mismatched training sets. {Boldfaced and underlined scores denote statistically significant improvements ($p < 0.05$) based on paired t-tests for related samples.} Among the unsupervised algorithms, \texttt{DEPSE-TL} consistently outperforms \texttt{DEPSE-IL} and UDiffSE+ in terms of SI-SDR, PESQ, and ESTOI, across almost all evaluation setups. For instance, \texttt{DEPSE-TL} achieves a notable SI-SDR improvement over UDiffSE+ when trained and evaluated on VB-DMD (13.87~dB vs. 10.40~dB); similarly, it outperforms UDiffSE+ on PESQ (2.17 vs. 2.05) and ESTOI (0.59 vs. 0.52) when trained on VB and tested on WSJ0-QUT. These results show that performance improves without a trade-off hyperparameter or likelihood approximation.

We also observe that \texttt{DEPSE-IL} performs comparably to or slightly better than UDiffSE+ across all settings. \texttt{DEPSE-TL}, on the other hand, exhibits more performance boost by providing a tractable likelihood. Noteworthy, the superiority of \texttt{DEPSE-TL} in SI-SDR does not imply better SI-SIR and SI-SAR. In fact, \texttt{DEPSE-TL} tends to 
introduce less artifacts (higher SI-SAR)  
than UDiffSE+ and \texttt{DEPSE-IL}, but retains slightly more residual interference (lower SI-SIR). Overall, the gain in SI-SAR outweighs the loss in SI-SIR, resulting in higher SI-SDR and thus improved perceptual quality and intelligibility, as reflected by the PESQ and ESTOI metrics. 

The supervised baselines, SGMSE+ and Conv-TasNet, achieve the highest performance across most metrics, except when trained on WSJ0-QUT and evaluated on VB-DMD, where \texttt{DEPSE-TL} outperforms them. When training and testing are conducted on different corpora, supervised models exhibit significant performance drops. For example, training on VB-DMD and evaluating on WSJ0-QUT leads to a 64.6\% SI-SDR decrease for SGMSE+ (8.53 $\rightarrow$ 3.02) and a 53.8\% drop for Conv-TasNet (12.63 $\rightarrow$ 5.83). In contrast, \texttt{DEPSE-TL} experiences an SI-SDR drop of 31.6\% (3.89 $\rightarrow$ 2.66), highlighting the higher robustness of \texttt{DEPSE-TL} to training-test domain mismatch. This robustness of unsupervised algorithms is more pronounced in Table~\ref{tab:vb}.

\vspace{-2mm}
\section{Conclusion}
We introduced two diffusion-based, unsupervised speech enhancement algorithms that sample from an explicitly derived conditional transition distribution. Unlike previous approaches, which rely on posterior-score guidance with approximated likelihoods and adjust only the mean of the prior and require a tuned hyperparameter, our first method computes both the conditional mean and variance without tuning. Our second method further introduces a forward diffusion over noisy speech, enabling exact likelihood computation. Experiments show that this method consistently outperforms prior unsupervised baselines in speech quality, intelligibility, and robustness under mismatched conditions. The results also demonstrate greater robustness to shifts between training and test conditions compared to supervised baselines. 

\bibliographystyle{IEEEtran}
\bibliography{mybib}

\end{document}